\documentclass[]{IEEEtran}

\usepackage{graphicx}
\usepackage[]{acronym}
\usepackage[bottom]{footmisc}
\usepackage{subfig}
\usepackage[backend=bibtex,sorting=none,style=ieee]{biblatex}

\acrodef{LHCb}{Large Hadron Collider beauty}
\acrodef{OMNeT++}{Objective Modular Network Testbed in C++}
\acrodef{DAQPIPE}{DAQ Protocol-Independent Performance Evaluator}
\acrodef{HPC}{High Performance Computing}
\acrodef{LHC}{Large Hadron Collider}
\acrodef{LS2}{Long Shutdown 2}
\acrodef{GbE}{Gigabit Ethernet}
\acrodef{EB}{Event Building}
\acrodef{RU}{Readout Unit}
\acrodef{FU}{Filter Unit}
\acrodef{EM}{Event Manager}
\acrodef{BU}{Builder Unit}
\acrodef{VL}{Virtual Lane}
\acrodef{FLIT}{Flow Control Unit}
\acrodef{NED}{Network Description}
\acrodef{HCA}{Host Channel Adapter}
\acrodef{PTP}{Precision Time Protocol}
\acrodef{TDR}{Technical Design Report}

\bibliography{IEEEabrv,bib}

\title{FLIT-level InfiniBand network simulations of the DAQ system of the LHCb experiment for Run-3}
\author{\IEEEauthorblockN{Tommaso Colombo\IEEEauthorrefmark{1}, Paolo Durante\IEEEauthorrefmark{1}, Domenico Galli\IEEEauthorrefmark{2}\IEEEauthorrefmark{3}, Matteo Manzali\IEEEauthorrefmark{3}, Umberto Marconi\IEEEauthorrefmark{3}, Niko Neufeld\IEEEauthorrefmark{1}, Flavio Pisani\IEEEauthorrefmark{1}\IEEEauthorrefmark{2}\IEEEauthorrefmark{3}, Rainer Schwemmer\IEEEauthorrefmark{1}, Sébastien Valat\IEEEauthorrefmark{1}}\\\IEEEauthorblockA{\IEEEauthorrefmark{1}CERN, Geneva, Switzerland\\}\IEEEauthorblockA{\IEEEauthorrefmark{2}Alma Mater Studiorum - Università di Bologna, Bologna, Italy\\}\IEEEauthorblockA{\IEEEauthorrefmark{3}Istituto Nazionale di Fisica Nucleare - INFN, Sez. di Bologna, Bologna, Italy\\Email flavio.pisani@cern.ch}}

\begin{document}
	\maketitle
	\begin{abstract}
	The \ac{LHCb} experiment is designed to study differences between particles and anti-particles as well as very rare decays in the charm and beauty sector at the \ac{LHC}. The detector will be upgraded in 2019 and a new trigger-less readout system will be implemented in order to significantly increase its efficiency and take advantage of the increased machine luminosity.
In the upgraded system, both event building and event filtering will be performed in software for all the data produced in every bunch-crossing of the \ac{LHC}. In order to transport the full data rate of 32 Tb/s we will use custom FPGA readout boards (PCIe40) and state of the art off-the-shelf network technologies. The full event building system will require around 500 nodes interconnected together.
From a networking point of view, event building traffic has an all-to-all pattern, therefore it tends to create high network congestion. In order to maximize the link utilization different techniques can be adopted in various areas like traffic shaping, network topology and routing optimization. The size of the system makes it very difficult to test at production scale, before the actual procurement. We resort therefore to network simulations as a powerful tool for finding the optimal configuration.
We will present an accurate low level description of an InfiniBand based network with event building like traffic. We will show comparison between simulated and real systems and how changes in the input parameters affect performances.
	\end{abstract}
	\section{Introduction}
		\begin{figure}[hbt]
			\centering
			\includegraphics[width=\columnwidth]{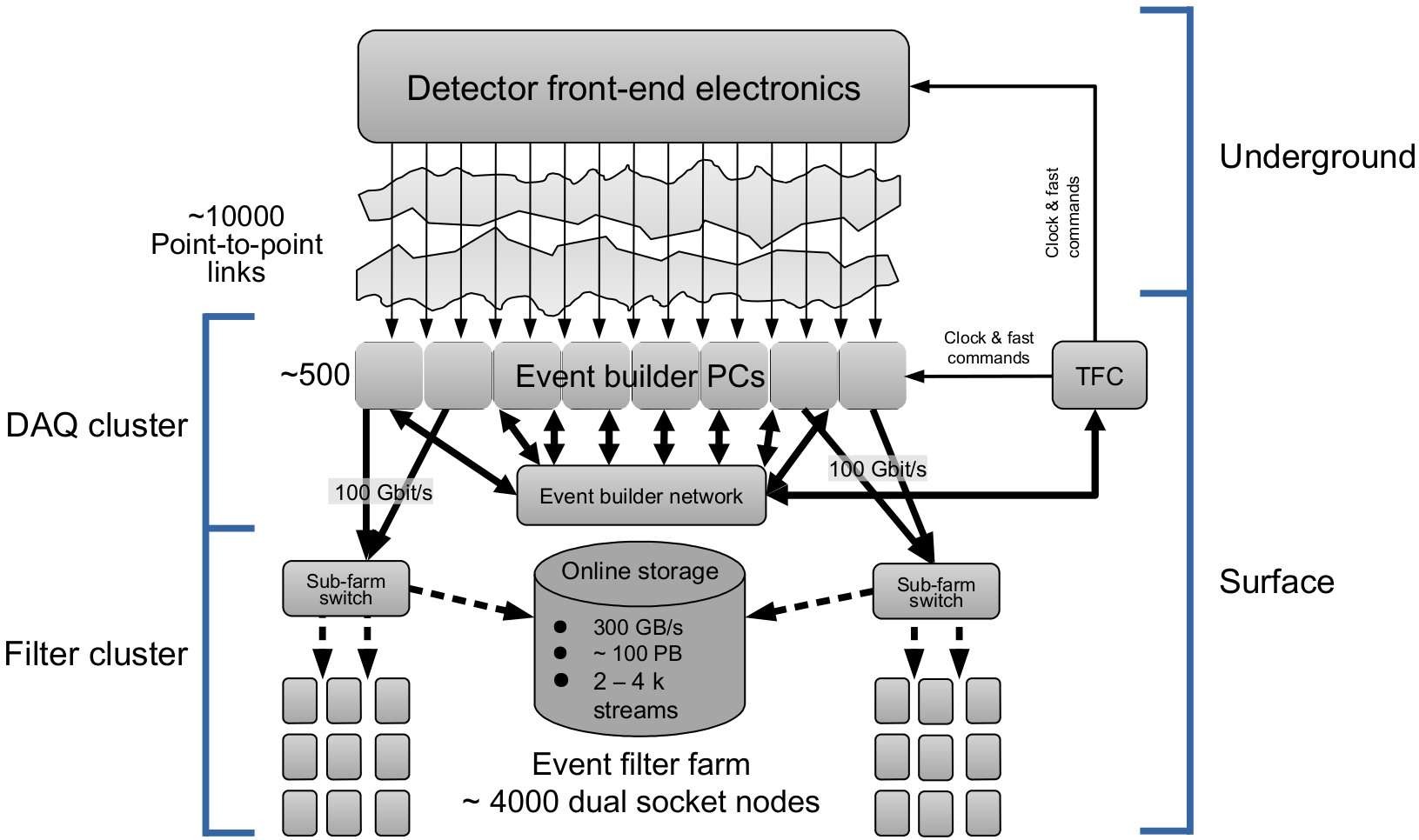}
			\caption{The architecture of the upgraded \ac{LHCb} readout system.\label{daq_system}}
		\end{figure}
	The \ac{LHCb} experiment\ \cite{LHCb} will receive a substantial upgrade\ \cite{tdr_upgrade} during the \ac{LS2} of the \ac{LHC}. One of the major changes during this upgrade process will be the installation of a completely new DAQ system without any low level hardware trigger, providing higher trigger yield at the luminosity foreseen after \ac{LS2}. To implement a trigger-less readout, the full bandwidth of \mbox{$\sim$32 Tb/s} produced by the detector must be forwarded by the event building network, in order to achieve this total throughput we are targetting a system composed of $\sim$500 nodes interconnected together using \mbox{100 Gb/s} networking technology, as shown in \figurename\ \ref{daq_system}.
	
	 In order to design and build a system with the above mentioned complexity we need extensive planning and testing, for this reason we developed \ac{DAQPIPE}. This software generates real event building traffic and can be configured in multiple ways in order to experiment with different network configurations and technologies. By only using \ac{DAQPIPE}, in order to test the scalability of the system, we need to access to \ac{HPC} clusters equipped with \mbox{100 Gb/s} capable interconnection networks. Because of the relative small number of suitable systems available in world, the waiting time can be very long and the network configuration may be suboptimal for event building tests.
	 
	 In this work, we present a low level simulation model that can be used, in parallel with tests  on real systems, to speed up the process of designing the event building network for a trigger-less readout system.
	 
	 \section{LHC\textup{b} Event Builder Architecture}
	 	In this section, we briefly  describe the DAQ's architecture of the \ac{LHCb} experiment for the Run-3 of the \ac{LHC}, because a full view is out of the scope of this paper we will focus on the network side of the system, a comprehensive view can be found in the \ac{TDR}\ \cite{tdr_upgrade}.
	 	\subsection{Event building architecture}
			\begin{figure}[hbt]
				\centering
				\includegraphics[width=\columnwidth]{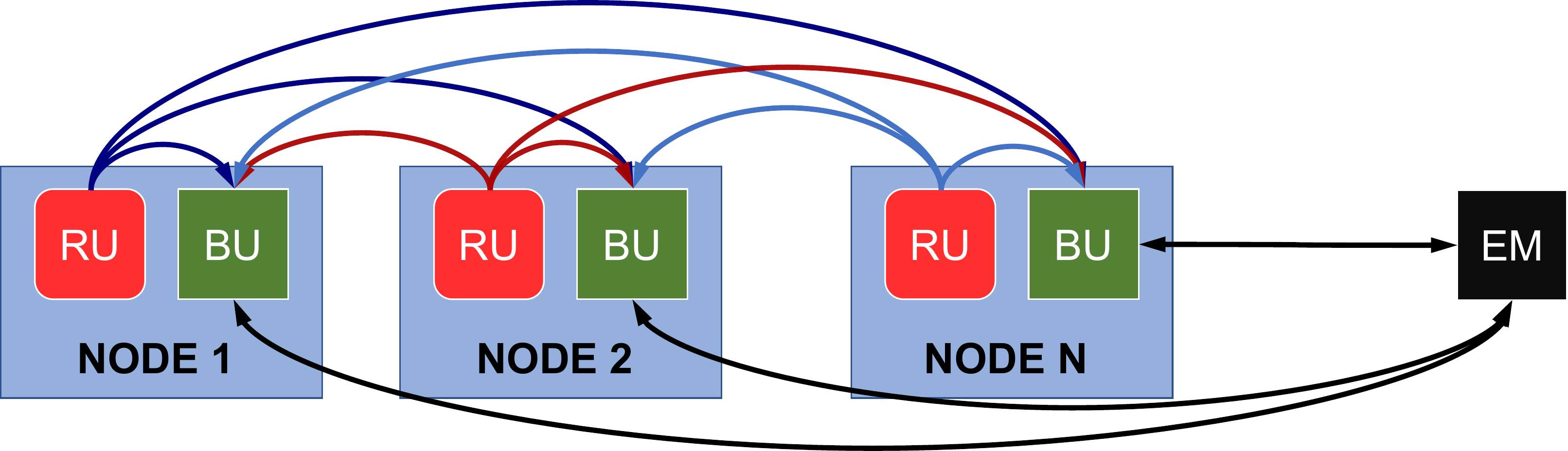}
				\caption{Event building architecture. The different arrows represent the multiple fragments gathered by the \acs{BU} while the black ones the control messages to and from the \acs{EM}.\label{eb_schema}}
			\end{figure}
	 		The \ac{LHCb} event building is composed of three main logical units:
	 		\begin{itemize}
	 			\item  \textbf{\ac{BU}} receives and aggregates the fragments into full events
	 			\item  \textbf{\ac{RU}} collects the fragments from the DAQ board and sends them to the \acp{BU}
	 			\item  \textbf{\ac{EM}} assigns which event is built on which \ac{BU}
	 		\end{itemize}
	 		As depicted in \figurename\ \ref{eb_schema} a \ac{BU} and a \ac{RU} are aggregated into one single node generating a 'folded' event builder, because the data traffic is always flowing from the \acp{RU} to the \acp{BU}, this architecture is used to fully exploit the full-duplex nature of the network and to reduce by a factor two the number of physical machines needed in the final system compared to a one-directional event builder.
	 		
	 		In the collective communication schema the traffic pattern of a folded event builder can be compared to an all-to-all with different data size for every fragment.
	 		
	 				In order to reduce the network congestion, generated by an all-to-all personalized exchange, we use the \textit{linear shifting} traffic scheduling technique, which can be explained as follows:
		\begin{itemize}
			\item We divide the all-to-all exchange into $N$ phases, where $N$ is the total number of nodes
			\item In every phase every node sends data to one destination and receives from one source
			\item During phase $n$ node $i$ sends to node $ (n+i)\%N $\footnote{The $\%$ symbol indicates the modulo operation}
		\end{itemize}
		If the aforementioned conditions are respected for all the phases then we have a linear shifting scheduling. In a real world scenario a mechanism for defining phases and synchronizing all the nodes must be provided.

	 	\subsection{Event building network}
	 	\begin{figure}[hbt]
				\centering
				\includegraphics[width=\columnwidth]{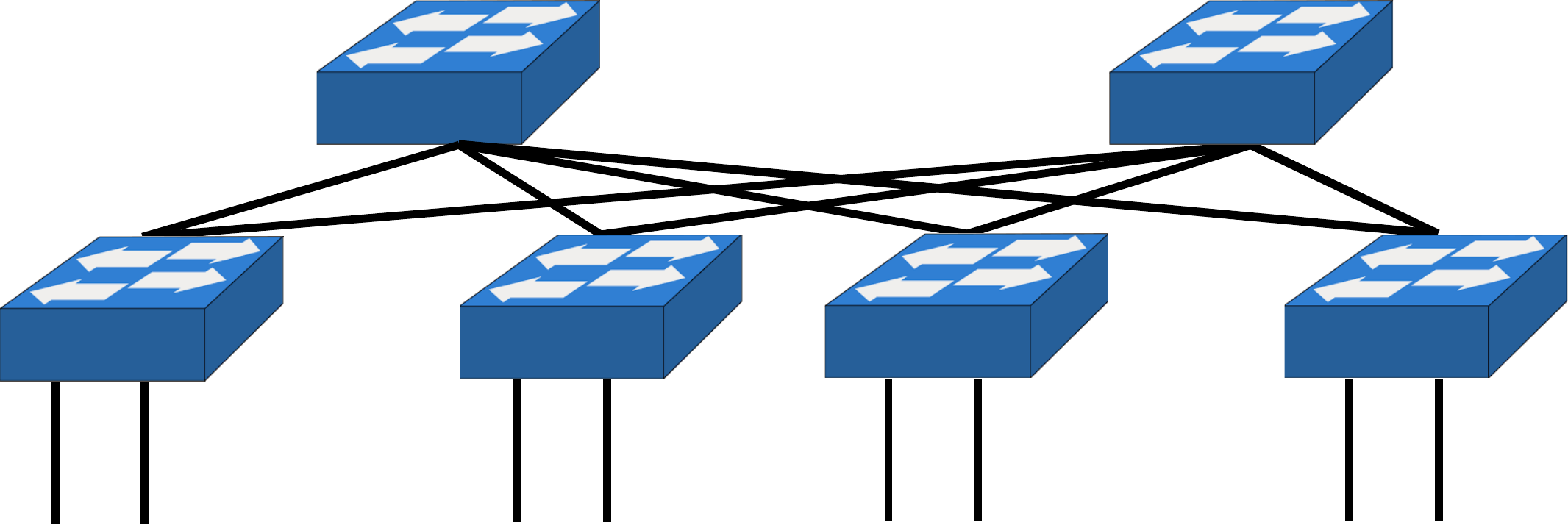}
				\caption{Fat-tree network build using switches with a radix of four. The two switches in the upper part are called spine switches, while the four in the lower part are called leaf switches.\label{fat-tree}}
			\end{figure}
	 		From the networking point of view, the event building traffic tends to create congestion and high link utilization among all the nodes, therefore the selected network topology has to be non-blocking and provide full bisection bandwidth.
	 		
	 		For the implementation of the \ac{LHCb} event building network, we decided to use a folded Clos network as the one depicted in \figurename\  \ref{fat-tree}; often referred  to as fat-tree\footnotemark. We selected this particular topology because: it fulfils the aforementioned requirements; it is widely adopted and it is supported by switch vendors. In particular, the OpenSM subnet manager used in InfiniBand-based networks provides optimized routing for fat-tree topologies\ \cite{ib_routing}. This algorithm uses a constant one-to-one correspondence between the spine switch selected and the switch port used by the destination node. This particular routing algorithm provides a conflict-free path for all the packets that are following a perfect linear shifter.
	 		
	 		\footnotetext{From a rigorous point of view the network topology shown in \figurename\  \ref{fat-tree} is a folded Clos network, nevertheless in the industry and data center world, it is frequently referred to fat-tree. Even if the network topologies are not exactly the same from this point on we will use the industry standard naming 'fat-tree' instead of 'folded Clos'.}
	 	\subsection{Event building benchmark: \ac{DAQPIPE}}
	 		\ac{DAQPIPE}\ \cite{daqpipev1_adam}\cite{daqpipev1_daniel}\cite{daqpipev1_flavio} is a small benchmark application to test  network fabrics for the future 
			\ac{LHCb} upgrade. It emulates an event builder based on a local area network and it supports multiple network technologies through different communication libraries like: MPI, LIBFABRIC, VERBS and PSM2.
			
			\ac{DAQPIPE} can be used either in a PUSH or PULL schema and it supports different traffic shaping strategies to reduce network congestion. Technologies and protocols can be mixed in a plug-and-play way.

			The software provides an implementation of all the logical blocks required by the \ac{LHCb} event building and emulates reading data from a real DAQ board connected to the detector. All the fragments of the same emulated event are then sent through the network using the desired communication library and protocol, and then aggregated into the \ac{BU} selected by the \ac{EM}.
			
			In order to take advantage of the available bandwidth and reduce the CPU overhead, \ac{DAQPIPE} sends multiple fragments of multiple events in parallel. The number of fragments in flight and the number of events processed in parallel can be tuned via two parameters:
			\begin{itemize}
				\item \textbf{Credits:} number of events processed in parallel by the \ac{BU}
				\item \textbf{Parallel sends:} number of fragments of the same event in flight
			\end{itemize}
			
			In order to reduce the traffic congestion \ac{DAQPIPE} provides a barrel shift-like traffic shaping, without enforcing strong synchronization among the nodes\footnote{There is a version of DAQPIPE with enforced timing but it will not be considered for the purpose of this paper}.
			
	 \section{Simulation Model}
		The simulations model we developed is implemented using the \ac{OMNeT++} framework\ \cite{omnetpp}; this discrete event simulator primarily targets network simulations and offers multiple tools that can be used to accomplish different tasks: from describing the network topology to gathering advanced statistics from the simulated design. In order to simulate the \ac{LHCb} DAQ system, we mainly need two components: an accurate description of the network and a precise modelling of the DAQ traffic.
		
		Mellanox technologies has already contributed to an  \ac{OMNeT++} based InfiniBand \ac{FLIT} level simulation model. 	This model supports: link level flow control, static lookup-table-based routing, arbitration between multiple \acp{VL}\footnote{A \ac{VL} is the InfiniBand implementation of a Virtual Channel\ \cite{Duato_book} - i.e. a set of multiple  flow control independent channels multiplexed on to the same physical one -}, packet generation and fragmentation and packet arbitration; however, it is not updated and does not support the \mbox{100 Gb/s} flavour of InfiniBand (i.e. EDR). Therefore we decided to expand the library capabilities to  fulfil our requirements and to make it as accurate as possible. In order to obtain a realistic model behaviour, we performed a fine tuning of the parameters using information collected from real hardware available in our test laboratory. In particular, we focused on: buffer sizes, network latency, link flow control, packet arbitration, latency and jitter of our entire software stack including PCIe communication overheads.
		 \subsection{Modules Description}
		 \begin{figure}[hbt]
			\centering
		 	\subfloat[Switch port implementation]{
				\includegraphics[width=0.439\columnwidth]{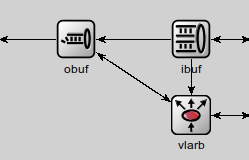}
				}
			 	\subfloat[Host implementation]{
				\includegraphics[width=0.45\columnwidth]{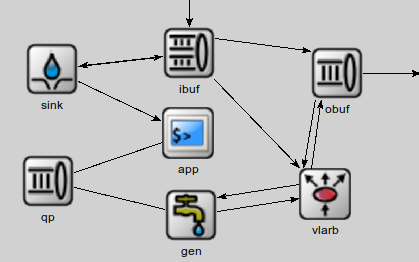}
				}
				\caption{Internal structure of a switch port and an host\label{modules}}
			\end{figure}
		 	\ac{OMNeT++} uses \textit{modules} as fundamental building blocks, hereinafter we provide a brief description of the main ones implemented:
		 	\begin{itemize}
		 		\item  \textbf{IBOutBuf:} buffer for outgoing \acp{FLIT}
		 		\item \textbf{IBInBuf:} buffer for incoming \acp{FLIT}
		 		\item  \textbf{IBVLArb:}  it implements arbitration among the different \acp{VL}
		 		\item \textbf{PktFwdIfc:} it provides destination ports to packets according to the static routing table
		 		\item \textbf{SwitchPort:} it combines input and output buffers with the \ac{VL} arbitration logic
		 		
		 		\item \textbf{IBApp:} it generates messages according to the selected traffic pattern 
		 		\item \textbf{IBWorkQueue:} queue for the different message coming from one or more applications
		 		\item \textbf{IBGenerator:} it arbitrates all the work queues and generates the packets and the \acp{FLIT} accordingly   
		 		\item \textbf{IBSink:} it receives the packets and notifies the IBApp module
		 	\end{itemize}
			
			\figurename\  \ref{modules} depicts how module can be interconnected together and generate more complex units.
		 \subsection{Topologies}
		 	
		 	In order to implement network topologies, \ac{OMNeT++} provides the \ac{NED} language which can be used to generate hierarchical and parametric networks. By using this powerful and flexible tool we implemented a parametric description of a fat-tree network. In view of analysing and comparing against real data collected on \ac{HPC} clusters we also implemented a Python script that generates \ac{NED} code by parsing the subnet manager information of the real cluster topology. In this way, we can study ideal topologies and compare them against real world systems with small imperfections like: missing nodes, swapped cables and suboptimal routing.

		 \subsection{Traffic injectors}
		Accurate traffic modelling is a key component for obtaining precise and realistic network simulation; in particular, in this work we used both synthetic and real application traffic. Our main target is to simulate the event building system of the \ac{LHCb} experiment, therefore a particular effort was put in an accurate replication of the \ac{DAQPIPE} traffic. Moreover we implemented two linear shifters with a different phase definition.
		A list and a brief description of the traffic injector implemented follows:
		\begin{itemize}
			\item \textbf{Fixed-size linear shifter:} it shifts destination after a fixed-size injection.
			\item \textbf{Time-window linear shifter:} it shifts destination after a fixed time interval. This injector uses a fixed grace period to absorb jitter, during this period the nodes are not allowed to send data, resulting in increased stability at the expense of a lower theoretical throughput.
			\item \textbf{\ac{DAQPIPE}:} an injector that replicates the real \ac{DAQPIPE} traffic. This traffic generator allows the user to change all the relevant parameters as in the real software.
		\end{itemize}	 
		 \section{Parameter Tuning}	
	 	The simulation model has several different parameters that need to be tuned and optimized to replicate the behaviour of real InfiniBand systems. For our event building studies we are interested in \mbox{100 Gb/s} networking solutions, therefore we tuned the model to replicate InfiniBand EDR hardware. In particular we use a Mellanox SB7700 EDR switch and Mellanox ConnectX-5 \acp{HCA}.
	 	
	 	Most of the basic parameters can be extracted from the InfiniBand architecture specification\ \cite{ib_spec}, e.g.: real bandwidth, header overhead, encoding overhead, link flow control behaviour, ecc. Advanced and hardware specific ones can be estimated performing real measurements and reverse engineering on the actual hardware.
	 	
	 	Crucial values for our simulations are: switch buffer size, link layer latency and PCIe latency.	 	
		\subsection{Switch buffer estimation}
			\begin{figure}[hbt]
				\centering
				\includegraphics[width=0.65\columnwidth]{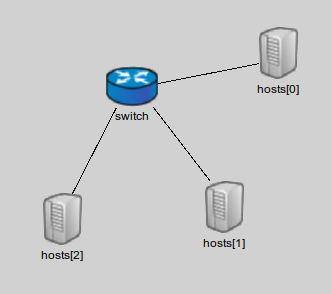}
				\caption{Setup used to generate congestion and estimate the switch buffer size. host0 sends at full speed data to host2, at the same time host1 sends packets of different sizes to create controlled congestion.\label{congestion}}
			\end{figure}
			In order to measure the switch buffer size we can use two different techniques\ \cite{ib_pack_anal}: analysing the link level flow control packets or generating congestion and monitoring the congestion indicator\footnote{The congestion indicator is the PortXmitWait counter which indicates the time, expressed in clock ticks, that a given port has been idling because of insufficient credits on the receiving buffer} on the various ports.
			
			 Decoding the information from the flow control packets produces a more accurate measure, but  it requires a low-level InfiniBand protocol analyser. Because there are no EDR capable protocol analysers available on the market, we decided to use the second strategy, and estimating the amount of buffering available in every switch port by measuring the performance counters.
			 
			 The setup used is depicted in \figurename\  \ref{congestion} and the procedure used to create congestion is the following:
			 \begin{itemize}
			 	\item host0 sends continuously to host2 at full speed
			 	\item host1 sends to host2 packets of increasing  size at regular intervals, to create congestion
			 	\item by reading the PortXmitWait counter and knowing the packet size we can estimate the buffer size of the switch
			 \end{itemize}
			
			Following this procedure we estimated a buffer size of \mbox{64 KiB} per port per \ac{VL} with 4 \acp{VL} enabled.
		\subsection{Link layer latency estimation}
			In order to measure the link layer latency, without using external protocol analysers, we decided to use the hardware timestamping feature of the IEEE 1588-2008 standard - i.e. \ac{PTP} - implementaion in the Mellanox \acp{HCA}.
			
			The path latency measure using \ac{PTP} produced an estimation of \mbox{170 ns} full delay using a direct attached copper cable, between two directly connected hosts.

		\subsection{PCIe latency modelling}
			\begin{figure}[hbt]
				\centering
				\includegraphics[width=\columnwidth]{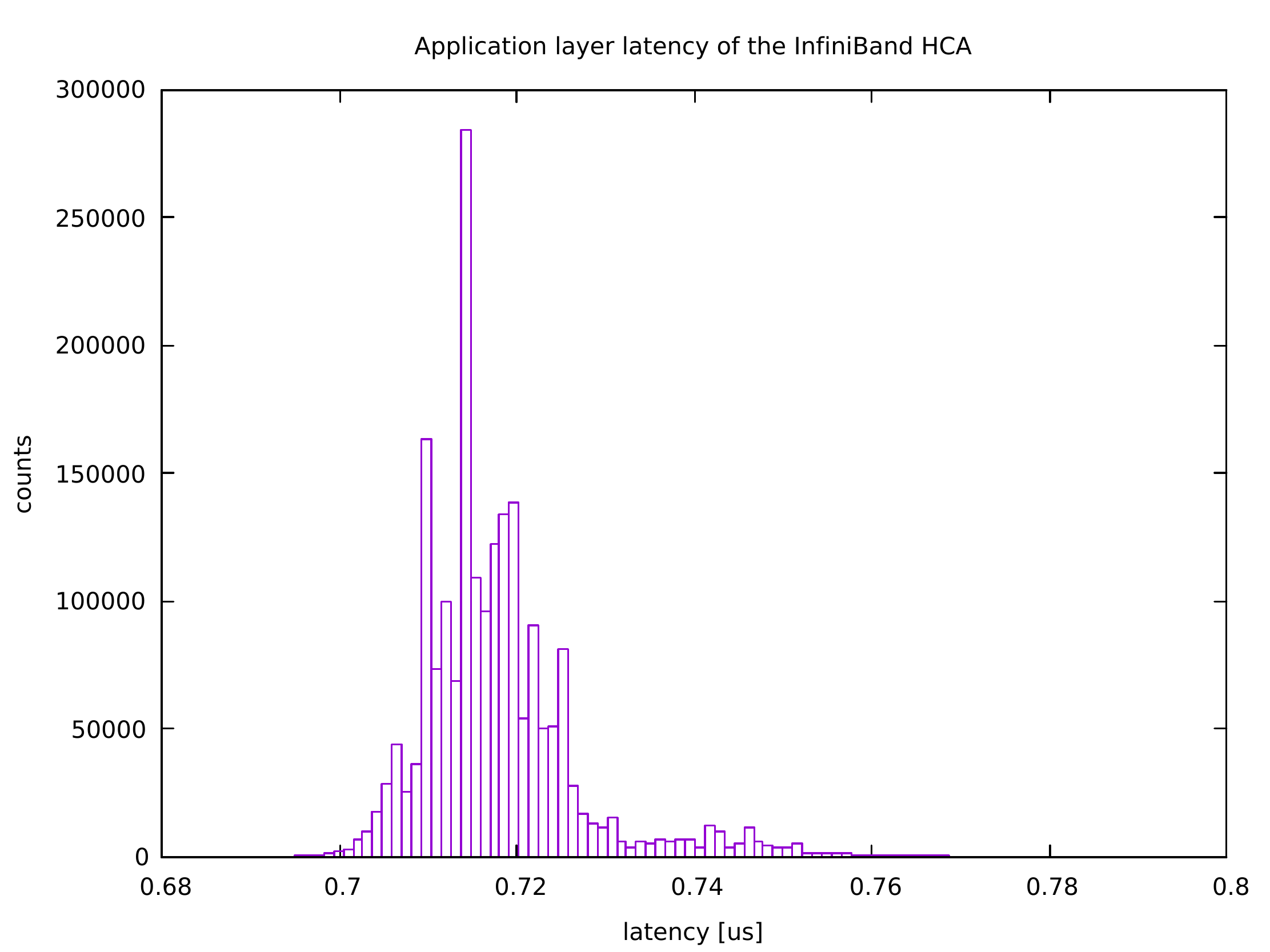}
				\caption{Application and PCIe latency of an InfiniBand EDR \ac{HCA}\label{lat}}
			\end{figure}
			The final piece in our model tuning is a realistic model PCIe and InfiniBand software stack latency; because of the non real time nature of modern computing systems and software, we decided decided to perform real world latency measures and replicate this behaviour in our simulation model.
			
			 The latency has been measured using the \textit{ib\_write\_lat} benchmark and subtracting the link layer latency, therefore this measurement include all the time needed from the hardware and software chain to make a packet available to the link layer.
			
			\figurename\  \ref{lat} shows the histogram of the latency measurements, the simulation model generates random number generates from this distribution to replicate latency and jitter of the real system.
			
	 \section{Results}
	 In this section we present some results obtained by simulating \ac{DAQPIPE} with the aforementioned simulation model. In particular we provide a comparison between the simulation and real data and a comparison of two different network topologies.
		\begin{figure}[hbt]
				\centering
				\includegraphics[width=\columnwidth]{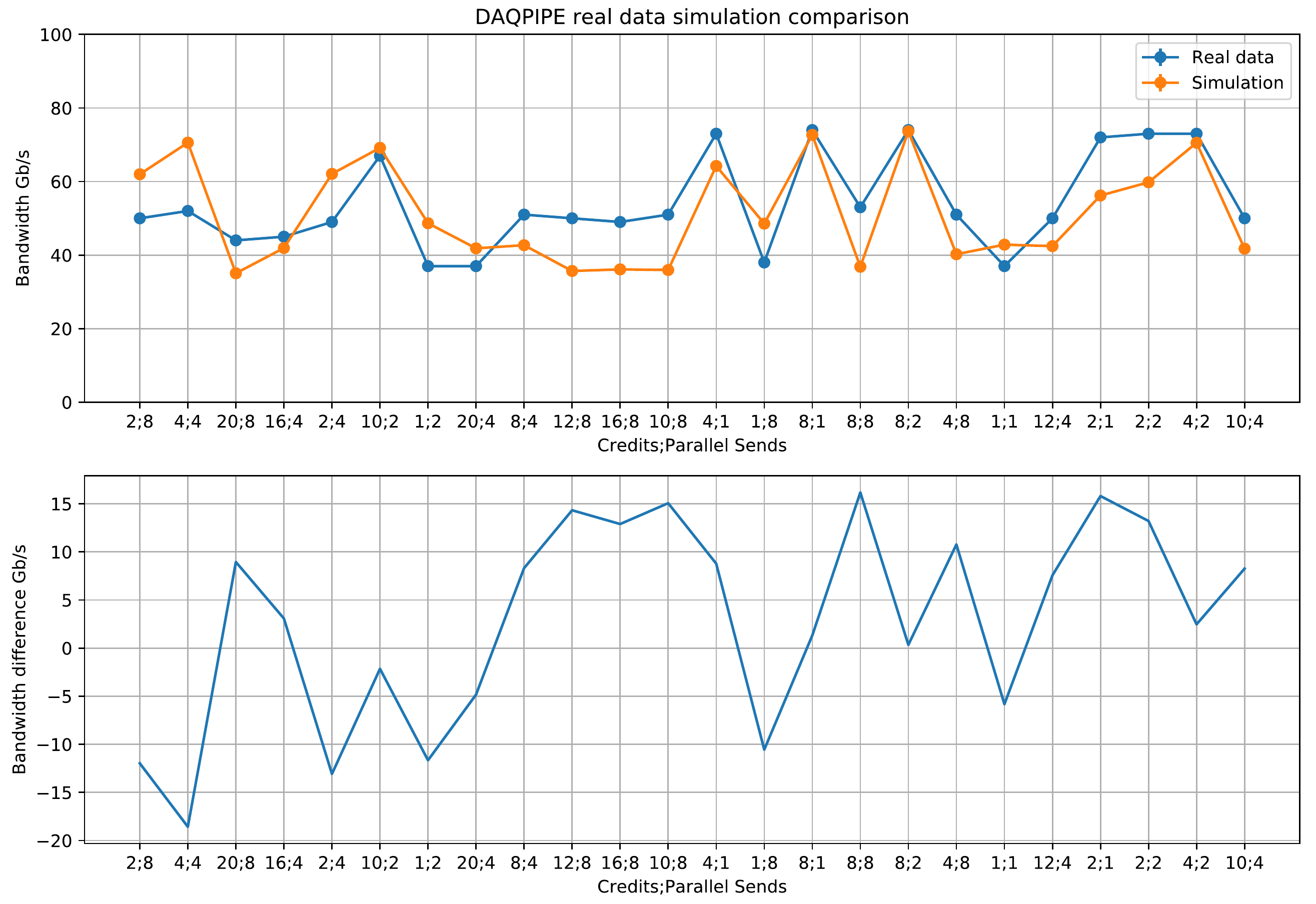}
				\caption{Comparison between real and simulated \ac{DAQPIPE} on a real \ac{HPC} cluster topology of 64 nodes\label{real_sim}}
			\end{figure}
		\figurename\  \ref{real_sim} shows a comparison between the simulated and the real \ac{DAQPIPE} for different values of the \textit{credits} and \textit{parallel sends} parameters.
		The real data are collected on an \ac{HPC} cluster of 64 nodes interconnected via a fat-tree-like network with: missing nodes, swapped cables and non-ideal routing. The simulation uses a replica of the same topology and the same routing of the real system.
		
		 From this plot we can confirm that the simulation can replicate the trend and the absolute value of the measurements performed on the real system.
			
		\begin{figure}[hbt]
				\centering
				\includegraphics[width=\columnwidth]{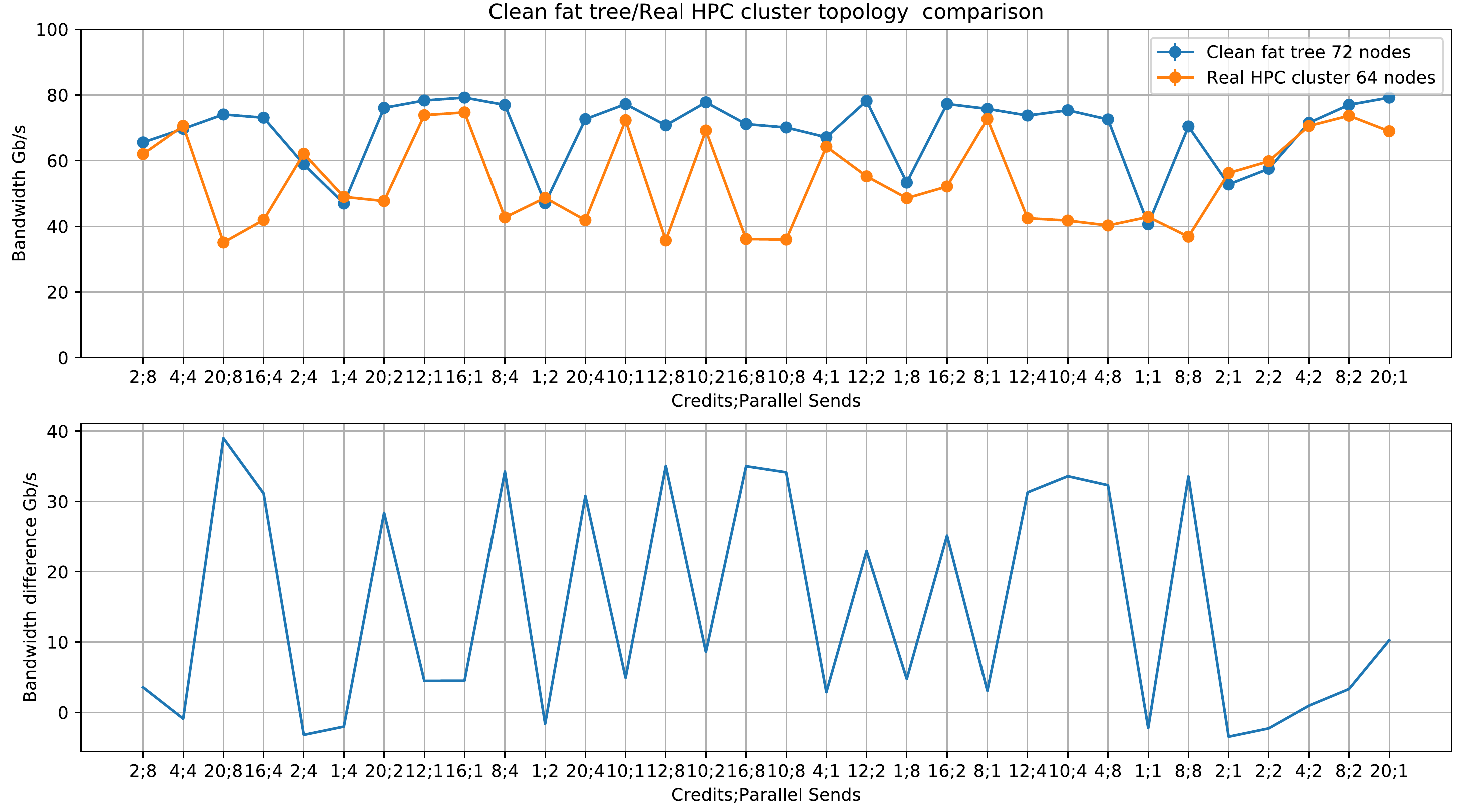}
				\caption{Comparison between simulated \ac{DAQPIPE} on a real \ac{HPC} cluster topology of 64 nodes and on a fat tree of 72 nodes\label{clean_cluster}}
			\end{figure}		
			
	In \figurename\  \ref{clean_cluster} we present a comparison of the performances of the simulated \ac{DAQPIPE} on two different topologies: a clean fat tree of 72 nodes and the real \ac{HPC} cluster of 64 nodes. 
	
	As we can see from the plot the performance loss is highly dependant on the parameters and can be as high as \mbox{40 Gb/s}, nevertheless the bandwidth drop for the best configuration is \mbox{$\sim$5 Gb/s}.
	
	 We can conclude that a non ideal topology affects the performances of \ac{DAQPIPE} and makes it more unstable, the performance drop can vary significantly and it is highly influenced by the configuration parameters and the topology itself.
	
	 \section{Conclusions and Future Work}
	 We have implemented an accurate low level model of our event building traffic based on the InfiniBand EDR fabric. We have tuned the model to achieve realistic results.
	 
	 We have validated our model against real data obtained on an \ac{HPC} cluster and we measured the impact of non ideal fat-tree topologies.
	 
	 We will run an extensive simulation campaign to evaluate the scalability of the system up to the required of \mbox{$\sim$500} nodes.
	 \printbibliography
\end{document}